\documentclass{kapedbk} 

\usepackage[dvips]{graphicx}
\upperandlowercase
\kluwerbib % will produce this kind of bibliography entry:

\def\apj#1{{\em Astrophys. J.} {\bf #1}}
\def\aj#1{{\em Astron. J.} {\bf #1}}
\def\mnras#1{{\em Mon. Not. R. astr. Soc.} {\bf #1}}
\def\aap#1{{\em Astron. Astrophys.} {\bf #1}}
\def\nat#1{{\em Nature} {\bf #1}}

\def\prl#1{{\it Phys. Rev. Lett.} {\bf #1}}
\def\araa#1{{\it Ann. Rev. Astron. Astrophys.} {\bf #1}}
\def\azh#1{{\it Astron. Zh.} {\bf #1}}
\def\apss#1{{\it Astrophys. Space Sc.} {\bf #1}}
\def\pasp#1{{\it Publ. ASP} {\bf #1}}
\def\etal{{et al.\/}\ }
\def\Mpc{$h^{-1}$~{\rm  Mpc}}

\begin{document}
\articletitle[Dark Matter]
{Dark Matter: Early Considerations}
%\author{Jaan Einasto, \altaffilmark{1}}
\author{Jaan Einasto \\
%\altaffiltext{1}{Tartu Observatory, 61602 T\~oravere, Estonia}
Tartu Observatory, 61602 T\~oravere, Estonia}

\begin{abstract}
A review of the study of dark matter is given, starting with earliest
studies and finishing with the establishment of the standard Cold Dark
Matter paradigm in mid 1980-s. Particular attention is given to the
collision of the classical and new paradigms concerning the matter
content of the Universe. Also the amount of baryonic matter, dark
matter and dark energy is discussed using modern estimates.

\end{abstract}


\section{Introduction}

Dark matter in the Universe can be described as the matter which has
practically zero luminosity and its presence can be detected only by
its gravity. Historically, the first modern study of the possible
presence of dark matter goes back to 1915, when \"Opik (1915)
determined the dynamical density of matter in our Galaxy in the Solar
vicinity.  The same problem was investigated by Oort (1932, 1960),
Kuzmin (1952a, 1955) and more recently by Bahcall (1985) and Gilmore,
Wyse \& Kuijken (1989).  Modern data suggest that there is little
evidence for the presence of a large amount of local dark matter in the
Solar vicinity. If some invisible matter is there, then it should be
in the form of brown dwarfs, jupiters or similar compact baryonic
objects.

A different type of dark matter is found around galaxies and in
clusters of galaxies. The first evidence for the presence of such
global dark matter was given by Zwicky (1933) from the dynamics of
galaxies in the Coma cluster.  The presence of dark matter in clusters
was questioned, and an alternative solution to explain large velocities of
galaxies in clusters was suggested by Ambartsumian (1958) -- the
instability of clusters of galaxies.  However, the evidence for
the presence of invisible matter in systems of galaxies accumulated,
first for our Local Group of galaxies (Kahn \&Woltjer  1959), and
thereafter  for all giant galaxies (Einasto, Kaasik \& Saar 1974,
Ostriker, Peebles \& Yahil 1974).  These results were questioned by
Burbidge (1975), Materne \& Tammann (1976). Independent
determination of rotation velocities of galaxies at large
galactocentric distances (Rubin, Ford \& Thonnard 1978, 1980) 
confirmed previous results on the presence of dark halos or coronas
around galaxies.  The nature of dark matter around galaxies
is not clear. Initially it was assumed that it consists of hot gas
(Kahn \& Woltjer  1959, Einasto 1974b). Modern data favour the
hypothesis that dark matter around galaxies is non-baryonic, either
neutrinos or some weakly interacting massive particles,
such as axions.  The neutrino-dominated dark matter is called hot, since
neutrinos move with very high velocities. The other type of dark
matter is called cold, as particle velocities are moderate.  The
cosmological model with cold dark matter (CDM) was suggested by
Blumenthal et al. (1984).  This model is presently accepted as the
standard. With the establishment of the cold dark matter concept
the early period of the study of dark matter was completed. 

Excellent reviews of the dark matter problem have been given by Faber
\& Gallagher (1979), Trimble (1987), Turner (1991) and Silk (1992),
alternatives to dark matter have been discussed by Sanders (1990).  In
this report I describe how astronomers developed step-by-step the
concept of dark matter. Such process is typical for the formation of a
new paradigm in our understanding of the Universe.  Particular
attention is given to the work on galactic modelling which has lead us
to the understanding of the structure of stellar populations and the
need for a new invisible population of dark matter in galaxies.  The
Power-Point version of the present report is available on the web-site
of Tartu Observatory, http://www.aai.ee/$\sim$einasto.

\section{Local Dark Matter}

Ernst \"Opik started his studies, being a student of the Moscow University.
One of the first problems he was curious about was the absorption of
light in the Galaxy and the possible presence of absorbing (invisible)
matter in it.  He developed a method to determine the density
of matter near the Galactic plane using vertical oscillations of
stars.  He concluded that there is no evidence for large amounts of
invisible matter near the Galactic plane (\"Opik 1915).

The dynamical density of matter in the Solar vicinity was investigated
again by Oort (1932), who arrived at a different answer. According to
his analysis the total density exceeds the density of visible stellar
populations by a factor of up to 2. This limit is often called the
Oort limit.  This result means that the amount of invisible matter in
the Solar vicinity could be approximately equal to the amount of
visible matter.

The work on galactic mass modelling in Tartu Observatory was continued
by Grigori Kuzmin.  He developed a new method for galactic mass
modelling using ellipsoids of variable density, and applied the theory
to the Andromeda galaxy (Kuzmin 1943), using the recently published
rotation data by Babcock (1939).  Next Kuzmin turned his attention to
our own Galaxy.  Here the central problem was the density of matter in
the Solar vicinity.  The mass density can be calculated from the
Poisson equation, where the dominating term is the derivative of the
gravitational potential in the vertical direction.  He found that this
derivative can be expressed through the ratio of dispersions of
velocities and coordinates in the vertical direction,
$C=\sigma_z/\zeta_z$; here $C$ is called the Kuzmin constant.  Kuzmin
(1952a, 1955) used data on the distribution of A and gK stars and
analysed the results obtained in earlier studies by Oort (1932) and
others.  He obtained a weighted mean value $C=
68$~km\,s$^{-1}$\,kpc$^{-1}$, which leads to the density estimate
$\rho = 0.08$~$M_{sun}$\ pc$^{-3}$, in good agreement with direct
density estimates of all known stellar populations (including
estimates for the mass in invisible low--mass stars and white
dwarfs). Two students of Kuzmin made independent analyses, using
different methods and observational data (Eelsalu 1959, J\~oeveer
1972, 1974) and confirmed Kuzmin results.

The local density problem was studied again by Hill (1960) and Oort
(1960); both obtained considerably higher local densities of 
matter, and argued that there exist large amounts of dark matter in
the Galactic disk.  More recently Bahcall (1984) constructed a new
multicomponent model of the Galaxy and determined the density of
matter in the Solar vicinity, in agreement with the Oort's (1932, 1960)
results.  The discrepancy between various determinations of the matter
density in the Solar vicinity was not solved until recently.  Modern data
have confirmed the results by Kuzmin and his collaborators (Gilmore,
Wyse \& Kuijken 1989).  Thus we came to the conclusion that {\em there
  is no evidence for the presence of large amounts of dark matter in
  the disk of the Galaxy}.  If there is some invisible matter near the
galactic plane, then it is probably baryonic (low--mass stars or
jupiters), since non-baryonic matter is dissipationless and cannot
form a highly flattened population.  Spherical distribution of the
local dark matter (in quantities suggested by Oort and Bahcall) is
excluded since in this case the total mass of the dark population
would be very large and would influence also the rotational velocity.

\section{Clusters and Groups of Galaxies}

The mass discrepancy in clusters of galaxies was found by Zwicky
(1933). He measured redshifts of galaxies in the Coma cluster and
found that the total mass of the cluster calculated from the velocity
dispersion using the virial theorem exceeds the sum of masses of
visible galaxies more than tenfolds. He concluded that the cluster
contains large amounts of invisible dark matter. 

For some reasons the work of Zwicky escaped the attention of the
astronomical community. The next step in the study of mass of systems
of galaxies was made by Kahn and Woltjer (1959).  They paid attention
to the fact that most galaxies have positive redshifts as a result of
the expansion of the Universe, only the Andromeda galaxy M31 has a
negative redshift of about 120 km/s. This fact can be explained, if
both galaxies, M31 and our Galaxy, form a physical system. A negative
radial velocity indicates that these galaxies have already passed the
apogalacticon of their relative orbit and are presently approaching
each other. From the approaching velocity, mutual distance and time
since passing the perigalacticon (taken equal to the present age of
the Universe) the authors calculated the total mass of the double
system. They found that $M_{tot} \geq 1.8 \times 10^{12}~ M_{sun}$.
The conventional mass of the Galaxy and M31 is of the order of $2
\times 10^{11}~ M_{sun}$, in other words, the authors found evidence
for the presence of additional mass in the Local Group of galaxies.
The authors suggested that the extra mass is probably in the form of
hot ionised gas; most of the paper was devoted to the analysis of the
physical state of the gas.  Using modern data Einasto \& Lynden-Bell
(1982) made a new estimate of the total mass of the Local Group, the
result was $4.5 \pm 0.5 \times 10^{12} M_{sun}$ for present age of the
Universe 14 Gyr. This estimate is in good agreement with new
determinations of total masses of M31 and the Galaxy including their
dark halos (see below).

The conventional approach for the mass determination of pairs and
groups of galaxies is statistical.  The method is based on the virial
theorem and is almost identical to the procedure used to calculate
masses of clusters of galaxies.  Instead of a single pair or group a
synthetic group is used consisting of a number of individual pairs or
groups. These determinations yield for the mass-to-luminosity (in blue
light) ratio the values $M/L_B = 1 \dots 20$ for spiral galaxy
dominated pairs and $M/L_B = 5 \dots 90$ for elliptical galaxy
dominated pairs (Page 1960, Burbidge \& Burbidge 1961, van den Bergh
1961, Karachentsev 1976, Faber \& Gallagher 1979).

The stability of clusters of galaxies was discussed in a special
meeting during the IAU General Assembly (Neyman, Page \& Scott 1961).
Here the hypothesis of Ambartsumian on the expansion of clusters was
discussed in detail. Van den Bergh (1961) drew attention to the fact
that the dominating population in elliptical galaxies is the bulge
consisting of old stars, indicating that cluster galaxies are old.  It
is very difficult to imagine how old cluster galaxies could form an
instable and expanding system.  These remarks did not find attention
and the problem of the age and stability of clusters remained open.

\section{Masses of Galaxies}

\subsection{Galactic Models}

The classical models of spiral galaxies were constructed using
rotation velocities. In contrast, the models of elliptical galaxies
were found from luminosity profiles and calibrated using central
velocity dispersions or motions of companion galaxies. An overview of
classical methods to construct models of galaxies is given by Perek
(1962).

Problems of the structure of galaxies were a major issue at the Tartu
Observatory since \"Opik's (1922) work on the distance of the M31,
where a simple hydrostatic model of this galaxy was constructed.  This
work was continued by Kuzmin who developed the major principles of
galactic modelling, and applied these to calculate models of M31 and
the Galaxy (Kuzmin 1943, 1952b, 1953, 1956a, b).  These were first
models with a continuous change of the spatial density (earlier sums
of ellipsoids of constant density were used).  However, individual
populations of galaxies were not represented in these models, in
contrast to the Schmidt (1956) model of the Galaxy where different
populations were included with ellipsoids of constant density. The
study of kinematic and physical properties of stellar populations was
made independently.  For a review of the early views on the structure
of galactic populations see Oort (1958), in Tartu this problem was
investigated by Rootsm\"ae (1961).

A natural generalisation of classical and Kuzmin models was the
explicit use of major stellar populations, such as the bulge, the
disk, and the halo, as well as the flat population in spiral galaxies
(consisting of young stars and interstellar gas). I did my PhD work on
stellar kinematics in 1955 and turned thereafter my interest to
galactic modelling.  My goal was twofold: first, to get more accurate
mass distributions in galaxies, and second, to find physical
parameters of main stellar populations in both spiral and elliptical
galaxies. My assumption was that similar stellar populations (say
bulges) in galaxies of different morphological type should have
similar physical parameters if their constituent stars have similar
age and metallicity distribution.  The methodical aspects of the new
multicomponent models were discussed in a series of papers in Tartu
Observatory Publications (in Russian with an English summary in
Einasto 1969a).  The spatial (or surface) density of practically all
stellar populations can be expressed by a generalised exponential law
(Einasto 1970b, 1974b, a similar expression has been used
independently elsewhere)
\begin{equation}
\rho(a) = \rho(0) \exp\left[-\left({a \over ka_0}\right)^{1/N}\right],
\end{equation}
where $\rho (0)=hM/(4\pi\epsilon a_0^3)$ is the central density, $a=
\sqrt{R^2+z^2/\epsilon^2}$ is the semi-major axis of the isodensity
ellipsoid, $a_0$ is the effective (mean) radius of the population, $h$
and $k$ are normalising constants, $M$ is the mass of the population,
$\epsilon$ is the axial ratio of isodensity ellipsoids, and $N$ is a
structural parameter, determining the shape of the density
profile. Here we assume that isodensity ellipsoids are concentric and
axially symmetric with a constant axial ratio for a given population.
The case $N=4$ corresponds to the de Vaucouleurs (1953) density law for
spheroidal populations (halo), $N=1$ corresponds to the classical
exponential density law, and $N=1/2$ to a Gaussian density law. The
practical procedure of the model construction is the following.
First, using photometric data for galaxies the structural parameters
$N$ of all major stellar populations are found.  Next, using
colorimetric and other data mass-to-luminosity ratios of populations
are derived.  Thereafter a preliminary mass distribution model is
found and the rotation (actually circular) velocity is calculated and
compared with observations.  From the difference of the calculated and
observed velocity corrections to model parameters are found.
Initially these corrections were found using a trial-and-error
procedure, later an automatic computer program was developed by our
young collaborator Urmas Haud (Einasto \& Haud 1989).

\subsection{Mass-to-luminosity Ratios and Models
  of Physical Evolution of Stellar Populations}

The method was applied to the Andromeda galaxy (Einasto 1969b, 1970a,
Einasto \& R\"ummel 1970a), and to our Galaxy (Einasto 1970b).  In the
case of the Andromeda galaxy the mass distribution model found from
the rotational data did not agree with the data on physical properties
of populations.  If we accepted the rotational velocities, based
mostly on radio observations (Roberts 1966), then the
mass-to-luminosity ratio, $M/L$, of central stellar populations
became very low, of the order of 1 in Solar units. On the other hand,
the spectral data (Spinrad 1966) suggested a much higher value, $M/L
\approx 17$.

The next problem was to find internally constitent values of physical
parameters of stellar populations of different age and composition.
For this purpose I developed a model of physical evolution of stellar
populations (Einasto 1971). When I started the modelling of physical
evolution of galaxies I was not aware of similar work by Beatrice
Tinsley (1968).  When my work was almost finished I had the
opportunity to read the PhD thesis by Beatrice. Both studies were
rather similar, in some aspects my model was a bit more accurate
(evolution was calculated as a continuous function of time whereas
Beatrice found it for steps of 1 Gyr, also some initial parameters
were different).  Both models used the evolutionary tracks of stars of
various composition (metallicity) and age, and the star formation rate
by Salpeter (1955). I accepted a low--mass limit of star formation,
$M_0 \approx 0.03~M_{sun}$, whereas Beatrice used a much lower mass
limit to get higher mass-to-luminosity ratio for elliptical
galaxies. My model yields a continuous sequence of population
parameters (colour, spectral energy distribution, $M/L$) as a function
of age.  The calculated parameters of stellar populations were
compared with observational data by Einasto \& Kaasik (1973).  The
available data supported relatively high values ($M/L\approx 10 - 30$)
for old metal-rich stellar populations near centres of galaxies;
moderate values ($M/L\approx 3 - 10$) for disks and bulges; and low
values ($M/L\approx 1 - 3$) for metal-poor halo-type populations.
Modern data yield slightly lower values, due to more accurate
measurements of velocity dispersions in central regions of galaxies,
and more accurate input data for models.

These calculations suggest that the rotation data by Roberts (1966)
are biased. To find the reason for this biasing, I analysed the
velocity field obtained from the radio observations. My analysis
suggested that low rotational velocities in the central regions are
due to a low spatial resolution of the radio beam (Einasto \& R\"ummel
1970b,c).  The corrected velocity field was in agreement with a higher
value of $M/L$ in the central regions of M31, suggested by direct
spectral data and models of physical evolution.

\subsection{Mass Discrepancy on the Periphery of Galaxies}

The second problem encountered in the modelling of M31 was the
rotation and density distribution on the periphery.  If the rotation
data were taken at face value, then it was impossible to represent the
rotational velocity with the sum of gravitational attractions by known
stellar populations. The local value of $M/L$ increases toward the
periphery of M31 very rapidly if the mass distribution is calculated
directly from the rotation velocity.  All known old metal-poor
halo-type stellar populations have a low $M/L \approx 1$; in
contrast, on the basis of rotation data we got $M/L > 1000$ on the
periphery of the galaxy, near the last point with a measured
rotational velocity.

There were two possibilities to solve this controversy: to accept the
presence of a new population with a very high $M/L$ (a very uncommon
property for an old stellar population), or to assume that on the
periphery of galaxies there exist non-circular motions.  We found
that the first alternative had several serious difficulties.  If the
hypothetical population is of stellar origin, it must be formed much
earlier than conventional populations, because all known stellar
populations form a continuous sequence of kinematical and physical
properties (Oort 1958, Rootsm\"ae 1961, Einasto 1974a), and there is
no place where to include this new population in this sequence.
Secondly, the star formation rate is proportional to the square of the
local density (Schmidt 1959, Einasto 1972), thus stars of this
population should have been formed during the contraction phase of the
formation of the population near its central more dense regions (where
the density is largest), and later expanded to the present distance.
The only source of energy for expansion is the contraction of other
stellar populations. The estimated total mass of the new population
exceeded the summed mass of all previously known populations.
Estimates of the energy needed for the expansion demonstrated that the
mass of the new population is so large that even the contraction of
all other stellar populations to a zero radius would not be sufficient
to expand the new population to its present size.  And, finally, it is
known that star formation is not an efficient process (usually in a
contracting gas cloud only about 1~\%~ of the mass is converted to
stars); thus we have a problem how to convert, in an early stage of
the evolution of the Universe, a large fraction of primordial gas into
this population of stars.  Taking into account all these difficulties
I accepted the second alternative -- the presence on non-circular
motions (Einasto 1969b), similar to many other astronomers (see
Materne \& Tammann 1976).  As I soon realised, this was a wrong
decision.

\subsection{Galactic Coronas}

In spring 1972 I was asked to give an invited review on Galactic
models at the First European Astronomy Meeting in Athens.  At this
time population models of galaxies had been calculated already for 5
galaxies of the Local Group and the giant elliptical galaxy M87 in the
Virgo cluster.  New rotation velocities suggested the presence of
almost flat rotation curves on the periphery of galaxies, thus it was
increasingly difficult to accept the previous concept of large
non--circular motions.  On the other hand, recently finished
calculations of the physical evolution of stellar populations
confirmed our previous view that it is extremely difficult to accept a
stellar origin of the hypothetical population.  My collaborator Enn
Saar suggested to abandon the idea that only stellar populations exist
in galaxies, to accept an idea that there may exist a population of
unknown nature and origin and to look which properties it should have
using available data on known stellar populations.  Quickly a second
set of models for galaxies was calculated, and parameters for the new
dark population were found.  To avoid confusion with the conventional
halo population I suggested to call the new population ``corona''
(Einasto 1974b).  The available data were insufficient to determine
the outer radii and masses of coronas. Rough estimates indicated that
in some galaxies the mass and radius of the corona may exceed
considerably the mass and radius of all stellar populations, taken
together.

To determine the parameters of galactic coronas more accurately
distant test bodies are needed. After some period of thinking I
realised how it is possible to check the presence of dark coronas
around galaxies.  If coronas are large enough, then in pairs of
galaxies the companion galaxies move inside the corona, and their
relative velocities can be used instead of galaxy rotation velocities
to find the distribution of mass around giant galaxies.  This test
showed that the radii and masses of galactic coronas exceeded the
radii and masses of parent galaxies by an order of magnitude!
Together with A. Kaasik and E. Saar we calculated new models of
galaxies including dark coronas.

In those years Soviet astronomers had the tradition to attend Caucasus
Winter Schools. Our results of galactic mass modelling were reported
in a Winter School in 1972. The next School was hold near the Elbrus
mountain in a winter resort, in January 1974. The bottom line of my
report was: {\em all giant galaxies have massive coronas, therefore
dark matter must be the dominating component in the whole universe (at
least 90~\% of all matter)}.  In the Winter School prominent Soviet
astrophysicists as Zeldovich, Shklovsky, Novikov and others
participated. In the discussion after the talk two questions
dominated: What is the physical nature of the dark matter?  and What
is its role in the evolution of the Universe?  A detailed report of
this study was sent to ``Nature'' (Einasto, Kaasik \& Saar 1974).

The need for massive halos had been already suggested by Ostriker and
Peebles (1973) to stabilise galaxies against bar formation. Soon after
our ``Nature'' paper Ostriker, Peebles and Yahil (1974) published similar
results using similar arguments. They used the conventional term
``halo'' for the dark population apparently not realising that this
population cannot be of stellar origin.

\subsection{Dark Matter Conferences 1975 }

The importance of dark matter for cosmological studies was
evident, thus Tartu astronomers organised in January 1975 a conference
in Tallinn devoted solely to dark matter. Historically this was the
first conference on dark matter. This conference is not well known, so
I give here the list of major talks:

%\begin{itemize}
%\item{} 
Zeldovich: ``Deuterium nucleosynthesis in the hot Universe and the
density of matter'';

%\item{} 
Einasto: ``Dynamical and morphological properties of galaxy systems'';

%\item{} 
Ozernoy: ``The theory of galaxy formation'';

%\item{} 
Zasov: ``The masses of spiral galaxies'';

%\item{} 
Fessenko: ``Difficulties of the study of dynamics of galaxy systems'';

%\item{} 
Novikov: ``The physical nature of galactic coronas'';

%\item{} 
Saar: ``Properties of stellar halos'';

%\item{} 
Doroshkevich: ``Problems of the origin of galaxies and galaxy
systems'';

%\item{} 
Komberg: ``Properties of the central regions of clusters of galaxies'';

%\item{} 
Vorontsov-Velyaminov: ``New data on fragmenting galaxies''.
%\end{itemize}

As we see, the emphasis of the conference was on the discussion of the
physical nature of dark matter and its role in the formation of
galaxies. These preliminary studies demonstrated that both suggested
models for coronas had difficulties. It is very difficult to explain
the physical properties of the stellar corona, also no fast-moving
stars as possible candidates for stellar coronas were found. 

Stellar origin of dark matter in clusters was disussed by Napier \&
Guthrie (1974); they find that this is possible if the initial mass
function of stars is strongly biased toward very low-mass stars.
Thorstensen \& Partridge (1974) discussed the suggestion made by
Cameron \& Truran (1971) that there may have been a pregalactic
generation of stars (called now population III), all of them more
massive than the Sun, which are now present as collapsed objects.
They conclude that the total mass of this population is negligible,
thus collapsed stars cannot make up the dark matter.

The gaseous corona of galaxies and clusters was discussed by Field
(1972), Silk (1974), Tarter \& Silk (1974) and Komberg \& Novikov
(1975).  The general conclusion from these studies is that coronas of
galaxies and clusters cannot consist of neutral gas (the intergalactic
hot gas would ionise the coronal gas), but a corona consisting of
ionised gas would be observable.  Modern data show that part of the
coronal matter in groups and clusters of galaxies consists of X-ray
emitting hot gas, but the amount of this gas is not sufficient to
explain flat rotation curves of galaxies.

The dark matter problem was discussed also during the Third European
Astronomical Meeting in summer 1975.  In contrast to the Tallinn
Meeting now the major dispute was between the supporters of the dark
matter concept and the older paradigm with conventional mass estimates
of galaxies.  The major arguments against the dark matter concept were
summarised by Materne \& Tammann (1976). They were as follows (see also
Burbidge 1975):

\begin{itemize}
\item{} The dark halo hypothesis is based on the assumption that
  companions are physical; if they are not then they do not measure
  the mass of the main galaxy, but characterise mean random velocities
  of galaxies;

\item{} Groups of galaxies are bound with conventional masses; the mean
  mass-to-luminosity ratios of groups are 4 and 30 for spiral and
  elliptical dominated groups, respectively;

\item{} The high masses of clusters may be explained by the high masses of
  the dominant cD galaxies; in other words -- there is no extra mass in
  clusters; 

\item{} Big Bang nucleosynthesis suggests a low-density Universe with
  the density parameter $\Omega \approx 0.05$; the smoothness of the
  Hubble flow  also favours a low-density Universe.
\end{itemize}

It was clear that by sole discussion only the presence and nature of
dark matter cannot be solved, new data and more detailed studies were
needed.

\subsection{Are Pairs of Galaxies Physical?}

In mid 1970s the main arguments for the presence of dark halos
(coronas) of galaxies and clusters of galaxies were statistical.  In
particular, the masses of double galaxies were determined by
statistical methods.  If companion galaxies used in mass determination
are not real physical companions but random interlopers, then the mean
velocity dispersion reflects random velocities of field galaxies and
no conclusions on the mass distribution around giant galaxies can been
made.

The difficulties connected with the statistical character of our
arguments were discussed already during the Caucasus Winter
School. Immediately after the school we started a study of properties
of companion galaxies.  The main question was: are companions true
members of the satellite systems, which surround giant galaxies.  Soon
we discovered that companion galaxies are segregated morphologically:
elliptical (non--gaseous) companions lie close to the primary galaxy
whereas spiral and irregular (gaseous) companions of the same
luminosity have larger distances from the primary galaxy; the distance
of the segregation line from the primary galaxy depends on the
luminosity of the primary galaxy (Einasto \etal 1974a).  This result
shows, first of all, that the companions are real members of these
systems -- random by-fliers cannot have such properties.  Second, this
result demonstrated that diffuse matter can have a certain role in the
evolution of galaxy systems.  The role of diffuse matter in galactic
coronas was discussed in detail by Chernin, Einasto \& Saar (1976).
Morphological properties of companion galaxies can be explained, if we
assume that at least part of the corona is gaseous.  On the other
hand, Komberg \& Novikov (1975) demonstrated that coronas cannot be
fully gaseous.  Thus the nature of coronas remained unclear.  Also we
found that dynamical and morphological properties of primary galaxies
are well correlated with properties of their companions (Einasto \etal
1976c).  Brighter galaxies have companions which move with larger
relative velocities than companions of fainter primaries.  A further
evidence of the large mass of the corona of our Galaxy came from the
study of the dynamics of the Magellanic Stream (Einasto \etal 1976a).

The status of the dark matter problem in galaxies was discussed during
the Commission 33 Meeting of the IAU General Assembly in Grenoble,
1976.  Here arguments for the presence of dark halos and its
non--stellar nature were again presented by Einasto, J\~oeveer \& Kaasik
(1976b).  But there remained two problems: 
\begin{itemize}
\item{} If the massive halo (or corona) is not stellar nor gaseous, of
what stuff is it  made of?

\item{} And a more general question: in Nature everything has its
purpose. If 90~\% of matter is dark, then this must have some
purpose. What is the purpose of dark matter?
\end{itemize}

\subsection{Additional Evidence for Dark Halos}

In mid 1970s Vera Rubin and her collaborators developed new sensitive
detectors to measure rotation curves of galaxies at very large
galactocentric distances.  Their results suggested that practically all
spiral galaxies have extended flat rotation curves (Rubin, Ford \&
Thonnard 1978, 1980, see also a review by Rubin 1987).  Now, for the
first time, it was possible to determine the mass distribution in
individual galaxies out to distances far superior to previous
data. The internal mass of galaxies rised with distance almost
linearly up to the last measured point (see Fig. 6 of Rubin et
al. 1978).  The concept of the presence of dark matter halos around
galaxies was confirmed with a high confidence.

Another very important measurement was made by Faber et al. (1977).
They measured the rotation velocity of the Sombrero galaxy, a S0
galaxy with a massive bulge and a very weak population of young stars
and gas clouds just outside the main body of the bulge.  Their data
yielded for the bulge a mass-to-luminosity ratio $M/L=3$, thus
confirming our previous estimates based on less accurate data, and
calculations of the physical evolution of galaxies.  Velocity
dispersion measurements of high accuracy also confirmed lower values
of mass-to-luminosity ratios of elliptical galaxies (Faber \& Jackson
1976).  These results showed that the mass-to-luminosity ratios of
stellar populations in spiral and elliptical galaxies are similar for
a given colour (the assumption used in our model calculations), and
the ratios are much lower than accepted in most earlier studies.

More recently the masses of clusters of galaxies have been determined
using the temperature of hot X-ray emission gas in clusters, and by
gravitational lensing.  These data are discussed in other reports
during this School.

By the end of 1970s most objections against the dark matter hypothesis
were rejected. In particular, luminous populations of galaxies have
found to have lower mass-to-luminosity ratio than expected previously,
thus the presence of extra dark matter both in galaxies and clusters
has been confirmed.  However, the nature of dark matter and its
purpose was not yet clear. Also it was not clear how to explain the
Big Bang nucleosynthesis constraint on the low density of matter, and
the smoothness of the Hubble flow.

\section{The Nature of Dark Matter}

\subsection{Neutrino-dominated Universe}

Already in 1970s suggestions have been made that some sort of
non-baryonic elementary particles may serve as candidates for dark
matter particles.  Gunn et al. (1978) considered heavy stable neutral
leptons as possible candidates for dark matter particles, however in a
later study Tremaine \& Gunn (1979) rejected this possibility. 
Cowsik \& McClelland (1973), Szalay \& Marx (1976) and Rees (1977)
noticed that neutrinos can be considered as dark matter particles; and
Chernin (1981) showed that, if dark matter is non-baryonic, then this
helps to explain the paradox of small temperature fluctuations of the
cosmic microwave background radiation.  Density perturbations of
non-baryonic dark matter start growing already during the
radiation-dominated era whereas the growth of baryonic matter is
damped by radiation.  If non-baryonic dark matter dominates
dynamically, the total density perturbations can have an amplitude of
the order $10^{-3}$ at the recombination epoch, which is needed for
the formation of the observed structure of the Universe.  This problem
was discussed in a conference in Tallinn in April 1981.  Here all
prominent Soviet cosmologists and particle physicists participated
(this conference was probably the birth of the astro--particle
physics).  The central problem was the nature of dark matter.  In the
conference banquet Zeldovich hold an enthusiastic speech: {\em
``Observers work hard in sleepless nights to collect data; theorists
interpret observations, are often in error, correct their errors and
try again; and there are only very rare moments of clarification.
Today it is one of such rare moments when we have a holy feeling of
understanding the secrets of Nature.''}  Non-baryonic dark matter is
needed to start structure formation early enough.  This example
illustrates well the attitude of theorists to new observational
discoveries -- the Eddington's test: {\em ``No experimental result
should be believed until confirmed by theory''} (cited after Turner
2000).  Now, finally, the presence of dark matter was accepted by
leading theorists.

The search of dark matter can be illustrated with the words of
Sherlock Holmes {\em ``When you have eliminated the impossible,
whatever remains, however improbable, must be the truth''} (cited by
Binney \& Tremaine 1987).

\subsection{Dark Matter and the Structure of the Universe}

After my talk at the Caucasus Winter School Zeldovich offered me
collaboration in the study of the universe.  He was developing a
theory of formation of galaxies (the pancake theory, Zeldovich 1970);
an alternative whirl theory was suggested by Ozernoy (1971), and a
third theory of hierarchical clustering by Peebles (1971).  Zeldovich
asked for our help in solving the question: Can we find some
observational evidence which can be used to discriminate between these
theories?

Initially we had no idea how we can help Zeldovich.  But soon we
remembered our previous experience in the study of galactic
populations: kinematical and structural properties of populations hold
the memory of their previous evolution and formation (Rootsm\"ae
1961, Eggen, Lynden--Bell \& Sandage 1962).  Random velocities of
galaxies are of the order of several hundred km/s, thus during the
whole lifetime of the Universe galaxies have moved from their place of
origin only by about 1~\Mpc\ (we use in this paper the Hubble constant
in the units of $H_0 = 100~h$ km~s$^{-1}$~Mpc$^{-1}$).  In other words
-- if there exist some regularities in the distribution of galaxies,
these regularities must reflect the conditions in the Universe during
the formation of galaxies.  Actually we had already some first
results: the study of companion galaxies had shown that dwarf galaxies
are located almost solely around giant galaxies and form together with
giant galaxies systems of galaxies.  In other words -- the formation
of galaxies occurs in larger units, not in isolation.

Thus we had a leading idea how to solve the problem of galaxy
formation: {\em We have to study the distribution of galaxies on
larger scales}. The three-dimensional distribution of galaxies, groups
and clusters of galaxies can be visualised using wedge-diagrams,
invented just when we started our study. My collaborator Mihkel
J\~oeveer prepared relatively thin wedge diagrams in sequence, and
plotted in the same diagram galaxies, as well as groups and clusters
of galaxies.  In these diagrams regularity was clearly seen: {\em
isolated galaxies and galaxy systems populated identical regions, and
the space between these regions was empty}. This picture was quite
similar to the distribution of test particles in a numerical
simulation of the evolution of the structure of the Universe prepared
by Doroshkevich et al. (1980) (preliminary results of this simulation
were available already in 1975). In this picture a system of high-
and low-density regions was seen: high-density regions form a
cellular network which surrounds large under-dense regions.
 
We reported our results (J\~oeveer \& Einasto 1978) at the IAU
symposium on Large-Scale Structure of the Universe in Tallinn 1977,
the first conference on this topic. The main results were: (1) galaxies,
groups and clusters of galaxies are not randomly distributed but form
chains, converging in superclusters; (2) the space between galaxy
chains contains almost no galaxies and forms holes (voids) of diameter
up to $\approx 70$~\Mpc; (3) the whole picture of the distribution of
galaxies and clusters resembles cells of a honeycomb, rather close to
the picture predicted by Zeldovich.  The presence of holes (voids) in
the distribution of galaxies was reported also by other groups: 
Tully \& Fisher (1978), Tifft \& Gregory (1978), and Tarenghi \etal
(1978) in the Local, Coma and Hercules superclusters, respectively.
Theoretical interpretation of the observed cellular structure was
discussed by Zeldovich (1978). 

Our analysis gave strong support to the Zeldovich pancake scenario.
This model was based essentially on the neutrino dominated dark matter
model.  However, some important differences between the model and
observations were detected.  First of all, there exists a rarefied
population of test particles in voids absent in real data. This was
the first indication for the presence of biasing in galaxy formation
-- there is primordial gas and dark matter in voids, but due to
low-density no galaxy formation takes place here (J\~oeveer, Einasto
\& Tago 1978, Einasto, J\~oeveer \& Saar 1980).  The second difference
lies in the structure of galaxy systems in high-density regions: in
the model large-scale structures (superclusters) have rather diffuse
forms, real superclusters consist of multiple intertwined filaments
(Zeldovich, Einasto \& Shandarin 1982, Oort 1983, see also Bond,
Kofman \& Pogosyan 1996).

\subsection{Cold Dark Matter}

The difficulties of the neutrino-dominated model became evident in
early 1980s.  A new scenario was suggested by Blumenthal, Pagels \&
Primack (1982), Bond, Szalay \& Turner (1982), and Peebles (1982);
here hypothetical particles like axions, gravitinos or photinos play
the role of dark matter.  Numerical simulations of structure evolution
for neutrino and axion-gravitino-photino-dominated universe were
made and analysed by Melott \etal (1983).  All quantitative
characteristics (the connectivity of the structure, the multiplicity
of galaxy systems, the correlation function) of this new model fit the
observational data well.  This model was called subsequently the Cold
Dark Matter (CDM) model, in contrast to the neutrino-based Hot Dark
Matter model.  Presently the CDM model with some modifications is the
most accepted model of the structure evolution. The properties of the
Cold Dark Matter model were analysed in detail in the classical paper
by Blumenthal \etal (1984).  With the acceptance of the CDM model the
modern period of the study of dark matter begins.

Numerical simulations made in the framework of the Cold Dark Matter
Universe (with and without the cosmological $\Lambda-$term) yield the
distribution of galaxies, clusters and superclusters in good agreement
with observations. These studies are too numerous to be cited here.
Also the evolution of the structure can be followed by comparison of
results of simulations at different epochs.  During the School a movie
was demonstrated showing the evolution of a central region of a
supercluster (the movie was prepared at the Astrophysical Institute in
Potsdam).  Here the growth of a rich cluster of galaxies at the center
of the supercluster could be followed. The cluster had many merger
events and has ``eaten'' all its nearby companions. During each merger
event the cluster suffers a slight shift of its position. As merger
galaxies come from all directions, the cluster sets more and more
accurately to the center of the gravitational well of the
supercluster. This explains the fact that very rich clusters have
almost no residual motion in respect to the smooth Hubble flow.
According to the old paradigm galaxies and clusters form by random
hierarchical clustering and could have slow motions only in a very
low-density universe (an argument against the presence of large amount
of dark matter by Materne \& Tammann 1976).

\subsection{The amount of dark matter}

In early papers on dark matter the total density due to visible and
dark matter was estimated to be 0.2 of the critical cosmological density
(Einasto, Kaasik \& Saar 1974, Ostriker, Peebles \& Yahil 1974). These
estimates were based on the dynamics of galaxies in groups and
clusters.  In subsequent years several new independent methods were
suggested.  A direct method is based on the distant supernova project,
which yields (for a spatially flat universe) $\Omega_m = 0.28 \pm
0.05$ (Perlmutter \etal 1998, Riess 1998). Here and below density
parameters are expressed in units of the critical cosmological
density.  Another method is based on
X-ray data on clusters of galaxies, which gives the fraction of gas in
clusters, $f_{gas} = \Omega_b/\Omega_m$.  If compared to the density
of the baryonic matter one gets the estimate of the total density,
$\Omega_m = 0.31 \pm 0.05 (h/0.65)^{-1/3}$ (Mohr \etal 2000). The
evolution of the cluster abundance with time also depends on the
density parameter (see Bahcall \etal 1999 for a review). This method
yields an estimate $\Omega_m = 0.4 \pm 0.1$ for the matter density.
The formal weighted mean of these independent estimates is $\Omega_m =
0.32 \pm 0.03$.  This density value is close to the value $\Omega_m =
0.3$, suggested by Ostriker \& Steinhardt (1995) as a concordant
model.  

More recently, the density parameter has been determined from
clustering in the 2-degree Field Redshift Survey (Peacock \etal 2001),
and from the angular power spectrum measurements of the cosmic
microwave background radiation with the Wilkinson Microwave Anisotropy
Probe (WMAP) (Spergel \etal 2003).  The most accurate estimates of
cosmological parameters are obtained using a combined analysis of the
Sloan Digital Sky Survey and the WMAP data (Tegmark \etal 2003).
According to this study the matter density parameter is $\Omega_m =
0.30 \pm 0.04$.  This method yields for the Hubble constant the value
$h = 0.70 \pm 0.04$ independent of other direct methods.  From the
same dataset the authors get for the density of baryonic matter, $h^2
\Omega_b = 0.0232 \pm 0.0012$, which gives $\Omega_b = 0.047$ for the
above value of the Hubble constant.  Comparing both density estimates
we get for the dark matter density $\Omega_{DM} = \Omega_m - \Omega_b
= 0.25$.

\section{Summary}

People often ask: Who discovered dark matter?  The dark matter story
is a typical scientific revolution (Kuhn 1970, Tremaine 1987).  As
often in a paradigm shift, there is no single discovery, the new
concept was developed step-by-step.

First of all, actually there are two dark matter problems -- the local
dark matter close to the plane of our Galaxy, and the global dark
matter surrounding galaxies and clusters of galaxies.  The milestones
of the local dark matter problem solution are the studies by \"Opik,
Oort, Kuzmin, Bahcall and Gilmore. Dark matter in the Galactic disk,
if present, must be baryonic (faint stars or jupiters). The amount of
local dark matter is low, it depends on the boundary between luminous
stars and faint invisible stars.

The story of the global dark matter also spans many decades.  It began
with the work by Zwicky (1933) on the Coma cluster, was continued with
the study by Kahn and Woltjer (1959) on the dynamics of the Galaxy-M31
system, and statistical determinations of masses and
mass-to-luminosity ratios of pairs, groups and clusters of galaxies.
For some reason, these studies did not awake the attention of the
astronomical community.  However, the awareness of the presence of a
controversy with masses of galaxies and galaxy systems slowly
increased.

Further development of the dark matter concept was influenced by
the East-West controversy during the Cold War (on this controversy see
Fairall 1998, p. 11 - 12).  The dark matter puzzle was solved in 1974
by two independent studies of masses of galaxies by Tartu and
Princeton astronomers. It was suggested that all giant galaxies are
surrounded by massive halos (coronas), and that dark matter is
dynamically dominant in the Universe.  As usual in scientific
revolutions, the general awareness of a crisis comes when the most
eminent scientists in the field begin to concentrate on the problem.
This happened when the Princeton group, Burbidge (1975) and Materne \&
Tammann (1976) published their contributions pro and contra the dark
matter hypothesis. In the following years experimenters devoted
themselves to finding new evidence in favour of (or against) the new
paradigm. The work by Rubin and collaborators on galaxy rotation
curves, our work on properties of satellite systems of galaxies and
the Magellanic stream, X-ray studies of clusters, as well as
investigation of gravitational lensing in clusters belong to this
type of studies.

The word on the crisis spread more rapidly in the East: the first dark
matter conference was held in Tallinn in 1975, the first official IAU
dark matter conference was held only ten years later.  The first popular
discussions of the problem were given in ``Priroda'' and ``Zemlya i
Vselennaya'' (the Russian counterparts of ``Scientific American'' and
``Sky \& Telescope'') by Einasto (1975) and Einasto, Chernin \&
J\~oeveer (1975), and also in the respective journal in Estonian. In
USA the first popular discussions were given many years later (Bok
1981, Rubin 1983). However, most experimental studies confirming the
dark matter hypothesis were made by US astronomers, and the cold dark
matter concept was also suggested by Western astronomers.

The new paradigm wins when its theoretical foundation is established. In
the case of the dark matter this was done by Blumenthal et al. (1984)
with the non-baryonic cold dark matter hypothesis. Also the need for
non-baryonic dark matter was clarified: otherwise the main
constituents of the universe -- galaxies, clusters and filamentary
superclusters -- cannot form.

In the following years  main attention was devoted to  detailed
elaboration of the concept of the cold dark matter dominated Universe.
Here a central issue was the amount of dark matter. Initially opinions
varied from a moderate density of the order of 0.2 critical density up to
the critical density. Only a few years ago it was clarified that dark
matter constitutes only 0.25 of the critical density, and the rest is
mostly dark energy, characterized by the cosmological constant or the
$\Omega_{\Lambda}$-term.  

To conclude we can say that the story of dark matter is not over yet
-- we still do not know of what non-baryonic particles the dark matter is
made of.

\begin{acknowledgments} 
I thank M. J\~oeveer and E. Saar for fruitful collaboration which has
lasted over 30 years.  This study was supported by the Estonian
Science Foundation grant 4695.
\end{acknowledgments}

\end{document}